# Characterizing Role Models in Software Practitioners' Career: An Interview Study

Mary Sánchez-Gordón
Department of Computer Science and Communication
Østfold University College
Halden Norway
mary.sanchez-gordon@hiof.no

Ricardo Colomo-Palacios
Escuela Técnica Superior de Ingenieros Informáticos Universidad Politécnica de Madrid
Madrid Spain
ricardo.colomo@upm.es

Alex Sanchez Gordon
Department of Innovation
LogicStudio
Panama Panama City
alex.sanchez@logicstudio.net

## ABSTRACT

A role model is a person who serves as an example for others to follow, especially in terms of values, behavior, achievements, and personal characteristics. In this paper, authors study how role models influence software practitioners' careers, an aspect not studied in the literature before. By means of this study, authors aim to understand if there are any salient role model archetypes and what characteristics are valued by participants in their role models. To do so, authors use a thematic coding approach to analyze the data collected from interviewing ten Latin American software practitioners. Findings reveal that role models were perceived as sources of knowledge, yet the majority of participants, regardless of their career stage, displayed a stronger interest in the human side and the moral values that their role models embodied. This study also shows that any practitioner can be viewed as a role model.

## 1 Introduction

Software's far-reaching influence has transformed the world around us. It has raised discussions about the need for a diverse and inclusive software engineering (SE) community [1,16]. Moreover, such a community has been called to prevent the replication of biases and the perpetuation of prevailing societal inequities [23]. A number of voices are underrepresented among software practitioners [21] and in response to this lack of diversity, many initiatives have emerged in the context of the software industry and academia.

"Big Tech" corporations [28–31] have diversity and inclusion (D&I) policies that serve as a reference for the world market. These initiatives involve actions such as giving voice to practitioners who are part of underrepresented groups and promoting role models. However, the term "role model" is loosely defined and thus subject to differing interpretations in both academic literature and everyday discourse [9,18]. Despite that fact, role modelling involves (1) one individual who perceives another individual as a role model and (2) the role model [12]. Unlike mentors, a relevant aspect of role modelling lies in the individual's cognitive acceptance of role models, outweighing the necessity of both an interactive relationship and the actions demonstrated by the role model [9]. This raises the question of how software practitioners define role models and, in turn, how role models influence software professionals' careers, an aspect not studied in SE literature previously (see section related works).

To fill this gap, this exploratory study aims to understand if there are any salient role model archetypes and what characteristics participants value in their role models. To do so, a qualitative research approach was conducted by using a thematic coding analysis of the interview data collected from ten Latin American software practitioners.

## 2 Related Works

Role models have received a considerable amount of attention in the Science, Technology, Engineering, and Mathematics (STEM) field, particularly in the context of education. For instance, Shin and London [24] conducted a study on STEM recruitment and retention issues in the USA that demonstrated the positive impact of role model biographies to foster increased STEM interest and identity among undergraduate students across various fields of study. A mobile app was proposed by the W-STEM project [8] to assist reference models in sharing their STEM-related stories with young girls. Fuesting et al. [7] exanimated that interactions with role models provide alternative pathways for communally oriented individuals to navigate STEM. A systematic literature review [11] on role models in STEM proposed four recommendations to implement role-model interventions based on the features of role models and students. Despite the efforts made, a recent scoping review [5] of interventions with role models found that evidence remains controversial regarding the efficacy of these interventions.

Grande [13] proposed a framework for role modelling in computing and engineering education. Moreover, de Souza et al. [6] explored how computer scientists can be used as role models to motivate students to continue pursuing a career in SE while Ouhbi et al.[19] explored the impact of incorporating a story about a role model in a SE course with the same purpose.

In the industrial context, a case study [14] on practice-based learning during the onboarding process at Google found that the concept of role models is explicitly presented as a good practice. Moreover, recent research [15,26] has found that a lack of role models is a key factor that discourages women from entering the tech or impedes their career progress. Trinkenreich et al. [26] found that some women practitioners have pointed out that this factor impacts the development of their self-confidence while van Breukelen et al. [2] identified that experienced women would like to be good role models. Beyond prior SE empirical research on women, studies on other underrepresented groups like LGBTQIA+ [6,27] and BIPOCs [20,22] (Black, Indigenous, and people of color) have highlighted the importance of role models.

Despite the valuable insights these previous studies provide, the first ones are focused on educational settings while the others do not primarily focus on role models. To the best of our knowledge, the potential impact of role models on the career construction of software practitioners has not been explored in the SE literature.

## 3 Research Approach

Our area of interest is behavioral software engineering. Particularly, this research focuses on understanding the software practitioners' perception of role models, and its possible impacts on their career development. Therefore, the following research questions were formulated:

**RQ1:** How do software practitioners define role models? Given that the concept of role models remains an ambiguous and imprecise term in the literature [18], we intend to understand it from the software practitioners' perspective.

**RQ2:** What characteristics do software practitioners value in their role models? We want to identify the characteristics that practitioners value. This is important to identify if archetypes of role models emerge through their career development.

### 3.1 Study Design

A qualitative interview study was conducted to answer these research questions. To get practitioners' perceptions, semi-structured interviews were used to collect data. This exploratory study has a sample size of ten participants. We used a purposive sampling approach to achieve gender balance between women and men practitioners while considering their career stages. The participants are part of our near network to make sure that their background is closely related to software development in an industrial setting. In addition, we chose to conduct this study in a software company since organizational culture and the interpretation of role-modelling behavior within that context can impact career aspirations [3]. It is also worth noting that this company is located in Panama.

The first author completed all the interviews to preserve consistency in the application of the method. The interview was conducted in Spanish, and the script was refined in a pilot study, with two interviewees that are not included in the analysis. Through this pilot study, we were able to identify questions that were ambiguous, repetitive, or difficult to understand. The interviews lasted from 30 to 45 minutes and each one was first voice-recorded and then transcribed to avoid the loss of data. In this way, it was easier to recall the content clearly and to gain a thorough insight into all the data. Each interview consisted of two parts.

At the beginning of the interview, we considered demographic information including job function. This allowed us to build rapport through small talk and a relaxed atmosphere between the interviewer and the interviewee. In the second part, we asked 8 questions about the participants' perception of role models and their experiences with them. For example: "What is a role model?" Do you think it is important to have role models? Why? The full interview script is available in [17].

**Ethics.** Before the interview, each participant was informed that participation is voluntary so that if they did not want to participate in the research, they could withdraw at any time, that the results would not be presented on their own, and that they were in a free environment to express their opinion. In addition, all participants were also informed that all data will be anonymized and kept confidential in this study. Verbal consent was obtained at enrollment, followed by obtaining written consent. This study was conducted in accordance with research ethics guidelines and regulations within the institutions and countries involved.

### 3.2 Data Collection

All interviews took place in October 2023 and were conducted through phone calls. With the participants' consent, all interviews were recorded and later transcribed into TXT files using Kahubi [32] before undergoing analysis. We invited 50 practitioners and all of them expressed their willingness to participate but nine were selected as participants based on their availability. In this exploratory study, we extended an invitation to an additional Peruvian woman employed in a bank located in the USA to check the findings related to gender identity and organizational culture.

### 3.3 Data Analysis

For data analysis, thematic analysis was selected since this study aims to identify, analyze, and report patterns within the data. Thematic analysis is particularly useful for providing a comprehensive description and organization of the interview data while facilitating the interpretation of various aspects relevant to the research topic [4]. Participants were assigned codenames ([P01] through [P10]) to anonymize the data.

First, we did an initial reading of the data (TXT files of audio transcription), and then we identified specific text segments and labeled them. Emerging codes were compared with earlier codes using constant comparison. In this way, the related codes were organized into themes to reduce overlap between codes. Finally, these themes were grouped to establish higher-order themes. During the analysis process, free-form notes were written to record thoughts and ideas about the codes. To facilitate this process, a qualitative data analysis platform called NVivo was used. All analyses were conducted in Spanish. and two authors discussed until achieving consensus about how the codes, themes and sub-themes were created. When conflicts were not resolved, the third author supported achieving consensus. Finally, we translated the codes and quotes into English.

## 4 Results and Discussion

In this section, we discuss the insights obtained by analyzing the data obtained from the interviews that contribute to addressing our research questions.

### 4.1 Demographic Overview

Table 1 summarizes the participants' demographic information about age, gender, years of experience, career stage and job. The average age is 31.4 with the following distribution 21-25 (2), 26-30 (3), 31-35 (2), 36-40 (1) 41-45 (2). Participants had a wide variety of software development career experience, ranging from 1.5 to 20 years, with an average of 8.75 years. Their combined expertise amounted to around 87 years.

**Table 1: Participant Demographics (P#: Participant #; Age; Gender; XT: Total experience (years); Career Stage; Job)**

| P# | Age | Gender | XT | Career Stage | Job |
|---|---|---|---|---|---|
| 01 | 36-40 | Male | 20 | Mid-career | Project Manager |
| 02 | 21-25 | Male | 1.5 | Exploration | Developer |
| 03 | 26-30 | Female | 4 | Establishment | Developer |
| 04 | 26-30 | Female | 1.5 | Establishment | UX Designer /Architect |
| 05 | 31-35 | Female | 10 | Establishment | Tester |
| 06 | 41-45 | Male | 20 | Mid-career | Developer |
| 07 | 31-35 | Female | 9 | Establishment | Tester |
| 08 | 41-45 | Female | 15 | Mid-career | Developer |
| 09 | 21-25 | Female | 1.5 | Exploration | Developer |
| 10 | 26-30 | Male | 5 | Establishment | Developer |

From a career stage perspective, participants were distributed across three phases, including the exploration stage (up to 25 years old, 2 individuals), the establishment stage (from 26 to 35 years, 5), and the mid-career stage (from 36 to 50 years, 3). [P04] has a relatively short amount of experience (1,5), primarily because of her career transition into software development. All participants were working on software projects and had different roles —Project Manager (1), Developers (6), Tester (2), UX Designer/Architect (1). Finally, participants were currently involved in projects across diverse domains, which included Transport, Finance and Banking, Retail, and Media and Communication. All participants are from Latin America, nine Panamanians and one Peruvian [P04].

### 4.1 Role Model Definition (RQ1)

A role model is a cognitive construct shaped by the context of the role aspirants. All participants —role aspirants— mentioned that they have self-selected their role models. For instance [P10] pointed out that *"a role model is someone in whom you see qualities [skills/competences] that you decide are either crucial or relevant to acquire ... or want to focus on acquiring in work or personal areas"*. We noticed that every participant provided a different definition, but all agreed that role models are important and have an impact on their personal and professional lives. One participant [P2] mentioned that *"a role model allows you to know where you want to go, where you want to get, and what you want to surpass"*. [P5] also said that *"a role model can be an example or a guide to follow and perform my job in a specific area"*. In simple words, *"a role model can be a life compass for different aspects"* [P01].

In addition, it is worth noticing that almost all participants needed some time to reflect on what a role model is and who their personal role models are. They appeared hesitant at some moments during the interview. This suggests that having a role model is not self-evident. Eventually, they mentioned more than one person as their role model since *"there is no one, there is no person who has 100% of everything. In fact, one can learn a little bit from each person"* [P01]. Thus, none of the participants had a single or global role model instead a pattern of multiple partial role models emerged. For instance, [P10] stated *"When you establish this reference point where you envision yourself, you amalgamate the various roles you observe"*. In other words, participants approach role models as a mosaic of characteristics from different people in their lives as mentioned in previous literature about medical education [25]. Such an approach is also in line with role model and career development literature [10].

Recurring elements that support role models' functions were found in this study. These findings are in line with the motivational theory of role modeling proposed by Morgenroth et al. [18]. Therefore, we grouped our findings using the schema proposed by Morgenroth et al., i.e., role models as behavioral models, representation of possible selves and inspirations.

**Behavioral model.** The behavioral value of role models focuses on observing and learning from software practitioners to help role aspirants become better professionals. Only two participants [P03, P05] did not focus on software professionals but they mentioned the importance of learning from others. In particular, [P01] said *"I have learnt a lot from him [his manager]"* and added, *"There was another project manager that I met on a project who I also learned from him because his personality was very similar"*. While [P05] claimed, *"I take delight in learning* *from the qualities of the person I consider a role model."* Likewise, [P6] stated *"I chose a model because they possess qualities or attributes that I would like to learn"* while [P10] claimed, *"One endeavors to acquire and sometimes even emulate these [role models'] ways of working"*. It suggests that software practitioners can serve as behavioral models when they reach the goal that the role aspirant is striving for, motivating them to progress towards an established objective. In fact, [P06] stated *"I see motivation"* whereas [P07] said, *"Those people [role models] have motivated me... because I want to make an effort"*. Both participants explained that the perception of goal embodiment plays a significant role in prompting vicarious learning.

**Representation of possible selves.** Role models can serve as representations of the possible and may also impact role aspirants' expectations of success. [P04] provided a good example for **changing self-stereotypes** by mentioning *"The technology field is somewhat challenging, isn't it? and even more so when you're a woman"* and then, she highlighted shared group membership and her role model success as follows *"she [role model] is one of those strong women you come across. They're not very common, in fact, there are very few that I've identified with. She's Peruvian, young and successful."* As representations of the possible, role models may also impact how role aspirants perceive external barriers.

For **changing perceived external barriers**, we identified that [P07] has emulated the behavior of her supervisor who is another woman. This role model shows that "it can be done", conveying to [P07] that gender is an obstacle that she can overcome and raising her expectations of success based on external factors. [P07] believe that *"she [her supervisor] is someone who inspires me to think "I can do it" ... and so ... I want to push myself"* and then added *"I want to be like her"*. Beyond inspiration, the role model overcame the barriers and achieved the objective and [P07] also believes that she can be like the role model in the future. These two women [P07, P04] provide similar insights regardless of their nationality, working country (Panama, the USA) and employer (software company, bank).

**Inspirations.** Most of the participants, except [P04, P06], explicitly mentioned inspiration during the interviews. For instance, [P02] highlighted that a role model is *"a person who inspires you to improve"* while [P03] stated that role models are important *"to get inspiration about what I could do"*. Likewise, seven participants, except [P02, P09, P10], used the word "admiration", e.g., [P04] said, *"A role model is someone I admire"* whereas [P06] noted that *"you can have as a role model someone you admire for how they lead their personal life, their family"*. However, a role model should be attainable as well [18]. [P03] illustrates this point by saying *"I do not say that they are my role models, rather I say that these characters [GitHub contributors] can inspire a lot when it comes to technology"*.

On the other hand, [P09] expressed *"My dad has given me many opportunities, so I must strive to achieve even greater things than he has"*, we identified this concept in seven interviews but not in three others [P03, P04, P10]. In this scenario, motivation seems to be triggered by processes such as personal identification, internalization, and admiration. Finally, most participants regardless of their gender pointed out that personal characteristics are not a constraint to choosing a role model. For instance, [P03] said *"Anyone can inspire anyone, right? It makes no difference what your age is"* and then she added *"... or gender ..."*

## 4.2 Role Model Characteristics (RQ2)

In this section, the findings are grouped into four themes that emerged from the interviews: Network, Influence, Time and Nature as shown in Table 2.

**Network.** There were near and distant role models from the role aspirants' network perspective. Personally known role models are near to the role aspirant. From the participants' perspective, the role models that have affected their professional life are relatives. In particular, parents were identified as a source of inspiration and motivation to work hard and persevere. Parent's socio-economic status also forced role aspirants to study hard to get a better life. Poverty in Latin America affects 32% of the total population, and our participants were not an exception. For instance, [P07] mentioned *"I've been working since I was 16 years old"* due to the need to ensure an adequate income for her family. Moreover, [P04] reflected on the impact of low socioeconomic background on privilege and opportunity by recalling *"There was not a university where I was born"* and *"my parents did not get an undergraduate degree"* and then reflected *"If you have a little more money, it is easier"*. [P04] also illustrated the importance of the differences between an aspirant role and their potential role models in generating expectations. It suggests that structural inequalities are involved in role modelling, and thus, further research is needed to explore intersectionality, and therefore the cultural context.

Besides relatives, former teachers were also mentioned [P2, P7]. Although [P03, P04, P06] did not mention relatives, the last two identified coworkers [P04, P06] and other professionals outside their company employer [P04]. In addition, six participants [P01, P04, P06, P07, P08, P10] highlighted the relevance of supervisors and/or managers as role models throughout their professional journey.

**Table 2: Role model characteristics by participant [P01-P10]**

| Role Model | 1 | 2 | 3 | 4 | 5 | 6 | 7 | 8 | 9 | 10 |
|---|---|---|---|---|---|---|---|---|---|---|
| **Network** | | | | | | | | | | |
| Near | • | • | | • | • | • | • | • | • | • |
| Distant | | | • | • | | | | | | • |
| **Influence** | | | | | | | | | | |
| Positive | • | • | • | • | • | • | • | • | • | • |
| Negative | | | • | | | | | • | • | • |
| **Time** | | | | | | | | | | |
| Permanent | • | • | | • | • | • | • | • | • | • |
| Transitory | | | • | | • | | | | • | |
| **Nature** | | | | | | | | | | |
| Real people | • | • | • | • | • | • | • | • | • | • |
| Characters | | | | | • | | | | • | • |

Distant role models were well-known figures in the technology area like Elon Musk, Mark Zuckerberg and Steve Jobs. The first two serve as inspiration to [P03] as she believes they promote communal opportunities. [P04] liked the ideas of Steve Jobs but she found issues with his personal life. Being aware of how to become like these figures or what it entails matters for most of our participants. For instance, [P10] claimed *"... at the end of the day ... [I think] we all want to be like Steve Jobs, we all want to be like Bill Gates, but I don't think it's a realistic thing"*. Participants in this study prefer to adopt personally known role models rather than distant ones.

**Influence.** While all participants mentioned positive role models, only four identified negative role models. [P10] notes there are *"role models that, in fact, are the opposite [of a positive role model]"*, he also reflected *"more than knowing what to do"* one can observe *"what not to do"*. In support of that, [P08] said *"I'm using them [negative role models] as an example of what I don't want to be"* while [P09] claimed, *"I'll try to avoid that [negative role model's] path"*. In addition, [P03] noted that *"If a role model's morality is questionable, one [role aspirant] can think in a very different way"* and reflected on values stating, *"It defines a lot of who you admire"*. It suggests that role aspirants can set approaching or avoidance goals. One unexpected finding was the relevance that participants placed on embodying moral values. In particular, participants mentioned their parents. One plausible explanation is their cultural background, but further research is needed. Participants in this study prefer positive role models rather than negative ones.

**Time.** There were permanent and transitory role models from the time perspective. [P10] declared *"some [role models] are replaced while others simply remain"*. All participants acknowledged that their parents are good examples of role models, so we named them as permanent role models. [P01] also provided another example by pointing out a manager with whom he has had a professional relationship for over a decade. The role modeling approach of three participants [P03, P05, P09] in this study unveils the opportunistic and transitory relevance of some people in their surroundings. Although [P03] did not mention relatives as role models and she did not remember the names of other near role models, she acknowledged the value of real people. On the contrary, [P05, P09] refrained from specifying names because they said that their approach was goal-oriented and focused more on professional journeys. However, they did not mention career development plans as [P04] explicitly did during her interview. Most participants in this study mentioned permanent role models but we identified that participants could be adopting transitory role models.

**Nature.** Role models could be real people or characters. [P09] claimed, *"I don't necessarily consider a person as a role model, rather my approach has always been based on the experiences, situations, and life journey I have observed"*. In addition, [P10] claimed *"a person can see a role model not so much in the author's book but in the book itself"* to refer to the character depicted within the narrative. [P05] was agreed with [P10]. In fact, [P10] did not remember the author's name but he was certain of the title's book. It suggests that, besides real people, a role model can be a character who may be entirely fictional or based on a real-life person. Those participants also mentioned social media platforms as ways to connect to people. Most participants in this study mentioned real people rather than characters.

### 4.3 Limitations

This study includes direct quotations from interviews to illustrate the interpretation of our participants' experiences and support the quality and credibility of our findings. As it was an exploratory study, the number of participants was limited to ten, threatening the findings in several ways. Although a purposive sampling method aimed to ensure diversity, we cannot claim the generalizability of our findings. In fact, we acknowledge that ten practitioners could not fully represent the experiences of Latin American software practitioners. Diversity in terms of career stage is another limitation related to lack of representativeness.

To improve the quality of the findings, further participants should be recruited. All the participants in this study were from Panamá, except one who was from Peru. Therefore, future research endeavors should broaden the scope of this study, for instance, by including Latin American practitioners working in the USA. or Europe. Researchers also acknowledge the cultural differences among Latin American countries. In addition, it would be interesting to conduct a worldwide survey to recruit more practitioners from diverse geographical locations, which is also relevant for role modelling due to the impact of cultural context. Finally, we also acknowledge that researchers may face challenges transferring our results to other cultural regions.

## 5 Conclusions

This exploratory study provides empirical evidence on the role modeling process perceived by Latin American software practitioners. By interviewing ten practitioners, we found that role modeling is a complex cognitive construct. Although having a role model is not self-evident, all participants acknowledge their relevance in their career development. Thematic analysis also shows that positive and near role models were the most common among the participants. However, they see role models as a mosaic of different characteristics from different real people or characters. An unexpected finding is that role models were seen as sources of knowledge, but most participants, regardless of their career stage, were more interested in the human side and moral values that they embodied.

Findings provide insights for future research endeavors. It would be interesting to explore the breakdown by different demographic factors or dismiss such a connection as not relevant. Furthermore, considering existing SE literature on mentoring would be worthwhile. We will expand our study to investigate how initiatives that promote role models are selecting them and how those role models learn to behave like role models. Finally, we hope that our findings help practitioners and researchers to better understand role modeling in our SE community.

## ACKNOWLEDGMENTS

Authors would like to thank all participants in this study and the anonymous reviewers of this article for their comments and constructive criticisms that led to important clarifications in the content of this version of the paper.